\documentclass[12pt]{article}
\usepackage{graphicx}

\def\eqref#1{eq.\ (\ref{#1})}
\def\figref#1{Fig.~\ref{#1}}
\def\dt{{\Delta t}}
\def\bz{{\bar{z}}}
\def\bzn{{\overline{z_n}}}
\def\avg#1{{\langle{#1}\rangle}}

\begin{document}

\title{Coarse-graining and self-similarity of price fluctuations}
\author{Yoshi Fujiwara$^1$ and Hirokazu Fujisaka$^2$\\
\\
$^1$  Keihanna Research Center,\\
  Communications Research Laboratory, Kyoto 619-0289, Japan\\
$^2$  Department of Applied Analysis and Complex Dynamical Systems,\\
  Graduate School of Informatics, Kyoto University,\\
  Kyoto 606-8501, Japan}
\date{\today}
\maketitle
\begin{abstract}
  We propose a new approach for analyzing price fluctuations in their
  strongly correlated regime ranging from minutes to months.  This is
  done by employing a self-similarity assumption for the magnitude of
  coarse-grained price fluctuation or volatility. The existence of a
  Cram\'er function, the characteristic function for self-similarity,
  is confirmed by analyzing real price data from a stock market.  We
  also discuss the close interrelation among our approach, the
  scaling-of-moments method and the multifractal approach for price
  fluctuations.
\end{abstract}

\vspace{2em}
% --------------------------------------------------

Scaling and self-similarity play an important role not only in physics
but also in various socio-economic systems including market
fluctuations, internet traffic, growth rates of firms and social
networks. If phenomenological theory is shown to be applicable to
observations, it would guide one to meaningful models and possibly to
universal and specific description of such systems without any
apparent ``Hamiltonian''.

Price fluctuations in speculative markets from physical viewpoint have
recently received great attention \cite{SELECTAE,IEP99,TFR00,IEEE99}.
Advances have been aided by concepts and techniques derived from
statistical physics and also by the new feasibility of analyzing
high-frequency financial data. Considerable statistical features have
been uncovered that, interestingly, are quite independent of asset
type (stock/foreign-exchange) and of markets (New York/Tokyo). In
fact, we shall describe below that price fluctuation is spectrally
white, but is not statistically independent since its magnitude
exhibits a strong correlation in the form of a power law
autocorrelation \cite{olsen93,liu99,gopi00} (see also references in
\cite{gopi00}).  Moreover, some have suggested that price fluctuations
might have similar properties in different time-scales, that is
partially supported by the multifractal analysis of price fluctuations
\cite{mandelbrot97,arnedo98,schmitt99}. In this paper we propose a new
approach to characterizing self-similarity in the strongly correlated
regime.  We actually show the existence of self-similarity and
characterize it both qualitatively and quantitatively from a new
statistical physical viewpoint. Specifically, we directly evaluate the
{\it invariant scaling\/} relation for fluctuations under {\it
  coarse-graining\/} of asset price returns.

Time-scale of our interest is related to the temporal features of the
price fluctuations known so far. Let $P(s)$ be the asset price at
time-step $s$, then the (logarithmic) return at $s$ for duration $t$
is defined by
\begin{equation}
  r_t(s)=\log(P(s)/P(s-t)) .
  \label{ret}
\end{equation}
Return is equal to the fractional change of price after duration $t$
when the change is small. We are interested in the
fluctuation $r_t$ for different values of $t$. Each transaction is
recorded for every stock in a particular market at the finest
time-scale. For the dataset from the New York Stock Exchange (NYSE)
used below, the typical time interval between successive transactions
was about 10 seconds for stocks with large (trading) volumes.

In a regime with time-scales of minutes and longer, considerable
statistical features are known to exist \cite{olsen93,liu99,gopi00}.
For an appropriate small time-scale $\dt$ (hereafter chosen as one
minute), (i) $r_\dt$ has an autocorrelation function (ACF) that almost
vanishes beyond several minutes of time lag, whereas (ii) $|r_\dt|$
has a long correlation ranging from several minutes to more than
several months or even a year. In fact, the ACF for $|r_\dt|$ has a
power-law decay that is valid for such a long range of time lags.
While (i) is interpretable as the absence of arbitrage using linear
prediction, the long memory (ii), known as volatility clustering in
economic literature, is not completely understood in its origin.

\begin{figure}
\begin{center}
  \includegraphics[width=14cm]{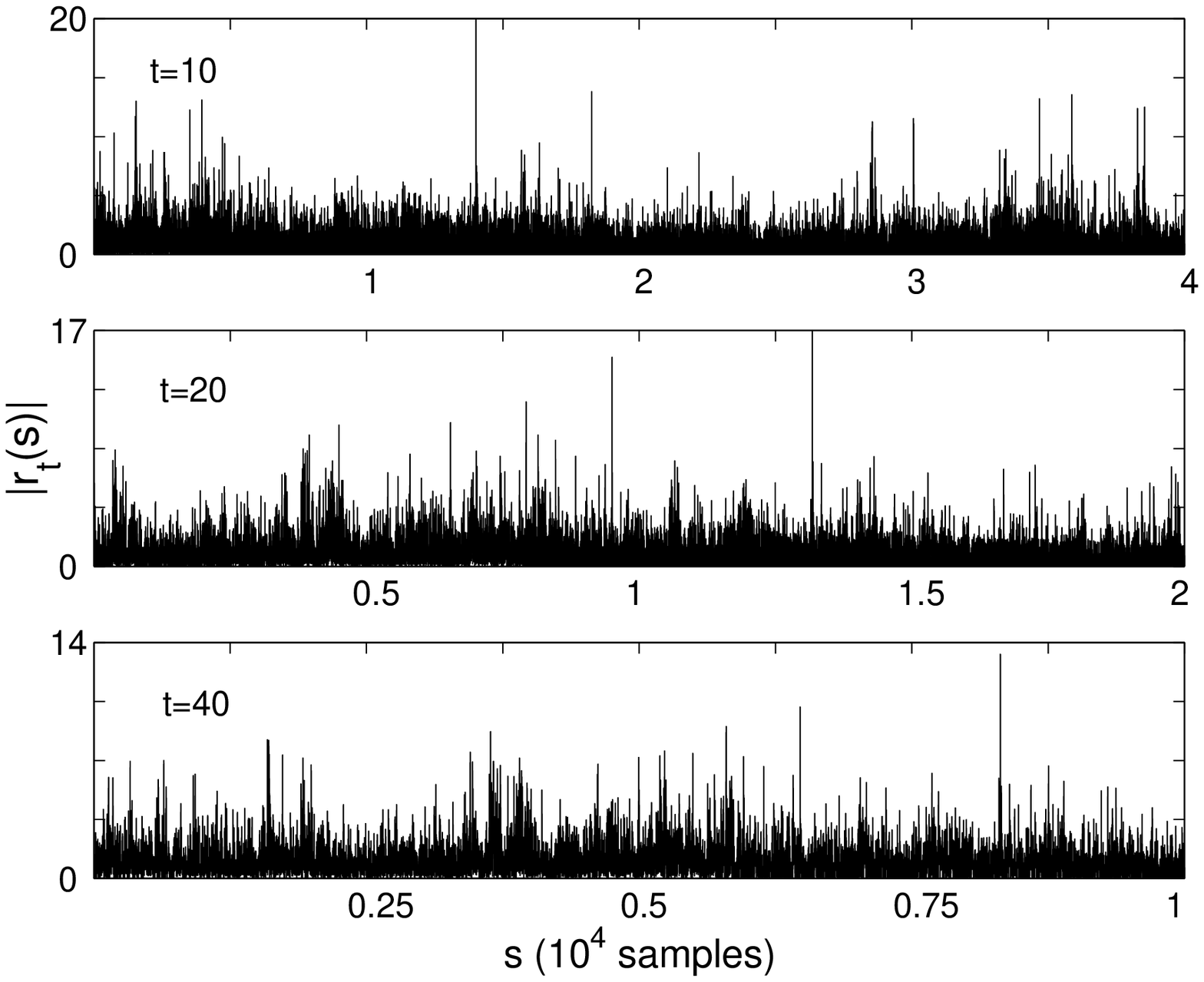}
\end{center}
\caption{%
  Time-series of price fluctuations magnitudes $y_t=|r_t(s)|$
  (normalized by standard deviation of each after removing intraday
  periodicity) for time-scales $t=10$, 20 and 40~min. Similar temporal
  evolutions are visible.}
\label{fig:ss}
\end{figure}

We assume from these observations that self-similarity property which
will be formulated shortly holds for time-scales between $t_s$ and
$T$, where $t_s$ is the small-, and $T$ the large-scale cut-off. The
value of $t_s$ is typically one minute or less, while $T$ may range
from several months to one year being regarded as a parameter. Using
an appropriate minimal time-scale $\dt$, we focus on the magnitude of
the coarse-grained return,
\begin{equation}
  y_t(s)\equiv
  \left|\sum_{k=0}^{t/\dt-1} r_\dt(s-k\dt)\right|
  =|r_t(s)| .
  \label{defy}
\end{equation}
The variable $t$ is the coarse-graining time-scale. \figref{fig:ss}
exemplifies in an actual data set the additional property of (iii)
self-similarity and intermittency for values of $y_t$ with different
time-scales $t$. The main aim of this work is to propose a new approach
for analyzing this self-similarity.

First, we introduce the coarse-graining step $n$ as
\begin{equation}
  t=t_n\equiv T\,e^{-n}
  \label{defn}
\end{equation}
for $n=0,1,\ldots,N$ ($N\equiv\log(T/t_s)\gg1$), and $y_n\equiv
y_{t_n}$. Next, we assume that the ratio $y_{n+1}/y_n$ is
statistically independent of $n$, i.e., statistically steady for $1\ll
n\ll N$. This {\it ansatz\/} is the expression of the self-similarity
assumption for price fluctuations. That is, by using the equation
\begin{equation}
  {y_{n+1}\over y_n}=e^{z_n} ,
  \label{ss}
\end{equation}
$z_n$ is statistically steady with respect to $n$ for the similarity
region $1\ll n\ll N$. If we introduce the exponent $\bz(t)$ as
\begin{equation}
  y_t\sim t^{-\bz(t)} ,
  \label{lexp}
\end{equation}
$\bz(t)$ corresponds to the local (H\"older) exponent in multifractal
theory.  Except for the $t$-dependence of the magnitude $y_t$, the
similar temporal evolution of $y_t$ for different values of $t$
(\figref{fig:ss}) is the origin of the self-similarity ansatz.

It follows immediately from \eqref{ss} that
\begin{equation}
  y_n=y_T\,\exp\sum_{k=0}^{n-1}z_k=y_T\,\exp(n\bzn) ,
  \label{yn}
\end{equation}
where $\bzn\equiv(1/n)\sum_{k=0}^{n-1}z_k$. We assume that
$y_T=y_{n=0}$ has no fluctuation in the sense that one can neglect
fluctuation if the coarse-graining scale is beyond the time-scale $T$.
Large deviation theory (e.g.\cite{FRISCH}) in probability theory tells
us that this self-similarity hypothesis, the statistical independence
of exponents $z_k$ in the coarse-graining step $k$, leads us to
express the probability density function (PDF) $Q_n(z)$ of the average
$\bzn$ as
\begin{equation}
  Q_n(z)\propto
  e^{-nS(z)} ,
  \label{qz}
\end{equation}
for $1\ll n\ll N$. Precisely speaking, \eqref{qz} is valid for $n\gg
n_c$, where $n_c$ is the correlation step of $z_n$ (e.g. defined as the
decay step of the correlation function of $\avg{\delta z_n\delta z_0}$,
where $\delta z_n\equiv z_n-\avg{z_n}$). Here the positive function
$S(z)$, being independent of $n$ for $n\gg n_c$, is known as a {\it
fluctuation spectrum\/} or {\it Cram\'er function}, whose functional
form characterizes the self-similar fluctuations and can be immediately
determined by observing the price fluctuations. Under the variable
transformation given by Eqs.\ (\ref{defn}) and (\ref{yn}),
\begin{equation}
  z_t(y)=\log(y/y_T)/\log(T/t)
  \label{zofy}
\end{equation}
(denoting $\bzn(y_n)$ as $z_t$ for short),
the PDF $Q_n(z)$ is transformed to the PDF of $y_t$, $P_t(y)$, as
\begin{equation}
  P_t(y)\propto
  {1\over y}\left({T\over t}\right)^{-S(z_t(y))} .
  \label{py}
\end{equation}

As a dataset for investigating this self-similarity hypothesis, we
used the Trade and Quote (TAQ) database of intraday transactions
provided by NYSE, Inc. The prices of a particular stock recorded
between Jan.~4, 1993 and June 30, 1999 were used. The median of the
time differences between quotes was 12~sec. The timing of the quotes
was used to construct a time-series of returns $r_\dt$ for
$\dt=$1~min, excluding midnights, i.e., the returns corresponding to
pairs of closing price and next day's open. The total dataset is
comprised of 638,901 records. Next we obtained an ensemble for $y_t$
with different time-scales $t$ by using \eqref{defy}. Other stocks
with high volumes were analyzed and yielded similar results to those
given below.

Removing the intraday periodicity from the time-series is a vital but
difficult task. We used essentially the same method as that in
\cite{liu99} (see also \cite{footnote1}). Note that a daily period for
the NYSE from open to close is approximately 390 minutes. The
observables adjusted with the intraday periodicity will be referred to
as $r_t$ and $y_t$.

\begin{figure}
\begin{center}
  \includegraphics[width=14cm]{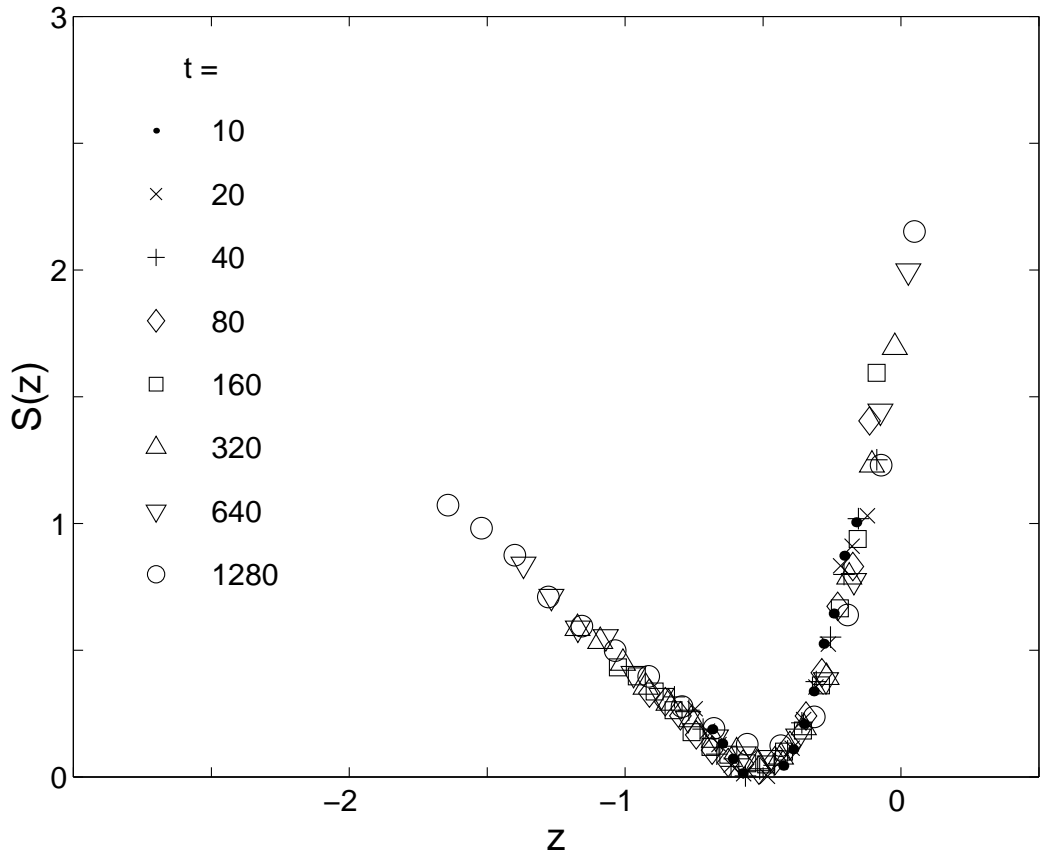}
\end{center}
\caption{%
  Variable $z$ vs. $S(z)$ (Eqs.\ (\protect\ref{zofy}) and
  (\protect\ref{scaling})) for time-scales $t=10,20,40,\ldots$ and
  1280~min, when $T=40960$ min. Plots lie on similar curve, which
  represents functional form of Cram\'er function $S(z)$. Note that
  $S(z)$ has minimum at $z\approx -0.5$.}
\label{fig:sz}
\end{figure}

Let us now calculate how the self-similarity ansatz formulated above
actually holds in the range of $t_s\ll t\ll T$. This
can be done by directly examining the scaling relation, \eqref{py},
rewritten as
\begin{equation}
  \log[y P_t(y)]/\log(T/t)= -S(z_t(y)) ,
  \label{scaling}
\end{equation}
where we ignored an additional constant term which corresponds to the
normalization of the PDF. The result is summarized in
\figref{fig:sz}. It shows that the curves for different coarse-graining
time-scales $t$ collapse. The value of $T$ is taken to be 40960~min,
$t=10,20,40,\ldots$ and 1280~min, and $y_T$ is set to be the mean of the
data for the cut-off time-scale $T$. The whole set of curves are shifted
vertically so that the minimums are at the origin of the vertical axis
(this corresponds to the additional constant ignored in
\eqref{scaling}). This invariant, i.e. the $t$-independent scaling
relation is the main result of the present work. It supports the
self-similarity ansatz and yields important information about the
functional form of $S(z)$. Note that the minimum $z_m$ for $S'(z_m)=0$
is approximately $-0.5$ demonstrating the diffusive nature of
$\avg{y_t}\sim\sqrt{t}$ in \eqref{lexp}.

It is easily shown that if the Cram\'er function is a linear function,
$S(z)=a(z-z_0)$, in a certain region of $z$, the PDF for $y_t$ takes
the power law $P_t(y)\sim y^{-a-1}$ in the corresponding region of
$y$. Data analysis shows that the region $dS(z)/dz<0$ can be closely
approximated by using $a=-1$. This corresponds to the fact that
$P_t(y)$ has a constant value for small fluctuations in $y$. It should
be noted that this region, and that for even smaller fluctuations, is
subject to the fact that price value is actually discrete (not shown
in \figref{fig:sz}), which is of limited interest. On the other hand,
the region $dS(z)/dz>0$ describes large fluctuations, which is related
to previous work on the scaling of moments reported in
Refs.~\cite{mandelbrot97,arnedo98,schmitt99}. That is, the moments
$\avg{y_t^q}$ of order $q$ scales with $t$ as $\avg{y_t^q}\propto
t^{\phi(q)}$. Such scaling is related to the approach presented here.
Actually, the relation between $\phi(q)$ and $S(z)$ can be found if we
note that
\begin{equation}
  \avg{y_t^q}\propto y_T^q\int_{-\infty}^\infty
  (T/t)^{qz-S(z)}dz ,
  \label{ytq}
\end{equation}
by using \eqref{py}. If we assume that the scaling is of the form,
$\avg{y_t^q}\propto t^{\phi(q)}$, and $S''(z)>0$, evaluating
\eqref{ytq} by using the steepest-descent method yields
\begin{equation}
  \phi(q)=\min_z[S(z)-qz] .
  \label{phi_sz}
\end{equation}
Thus $\phi(q)$ and $S(z)$ are related to each other by the Legendre
transformation of \eqref{phi_sz}.

Our self-similarity formulation originates from Kolmogorov's lognormal
theory in the statistical theory of turbulence\cite{K62}, and has been
recently applied, from the statistical viewpoint of large deviation
theory, to analyze the shell model of turbulence \cite{watanabe00} and
the long-time correlation in on-off intermittency
\cite{fujisaka00}. Indeed the return $r_t$ for duration $t$
corresponds to the velocity difference across a spatial distance
arbitrarily chosen in the inertial subrange in turbulence, where the
self-similar energy cascade process is observed \cite{K62,FRISCH}. In
the study of on-off intermittency which is typically observed right
after the breakdown of synchronization of coupled chaos
\cite{fujisaka00}, the difference of dynamical variables of chaotic
elements also shows self-similar characteristics. In connection with
this phenomenon, $r_t$ corresponds to the difference of dynamical
variables which exhibits on-off intermittency. In the latter analogy, 
we note that the corresponding variable possesses similar
properties as the above (i)--(iii) for the price fluctuations. The
present analysis is useful for these systems which show
self-similarity due to strongly correlated fluctuations either in
space or in time.

\begin{figure}
\begin{center}
  \includegraphics[width=14cm]{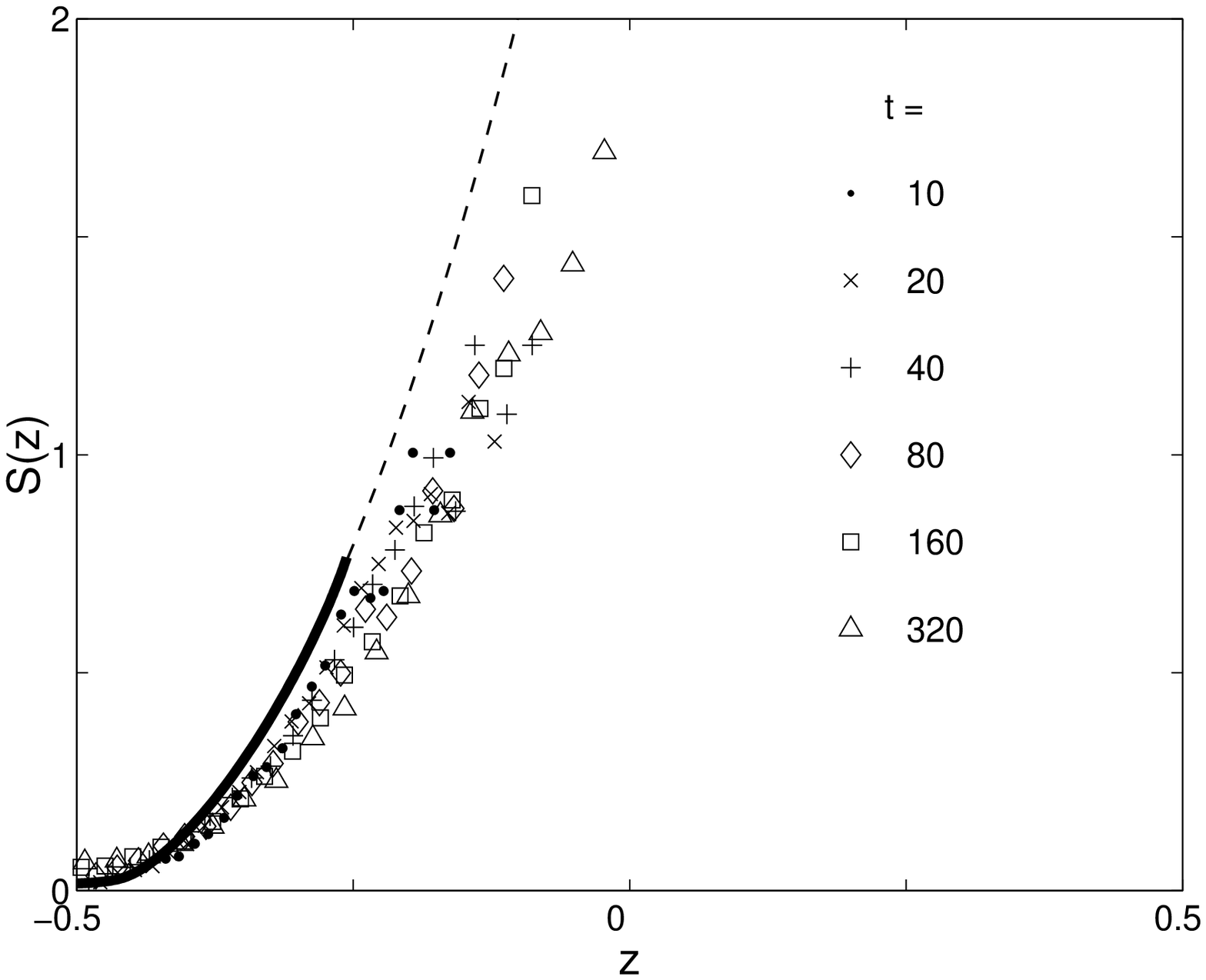}
\end{center}
\caption{%
  Details of Fig.\ \protect\ref{fig:sz} in region
  $dS(z)/dz\geq0$ for time-scales up to $t=320$ min. Compare with the
  result (heavy line) obtained from scaling of moments (of order $0\leq
  q\leq8$) by Legendre transformation. Broken line is
  quadratic-polynomial fitting for solid line and an
  extrapolation (lognormal approximation).}
\label{fig:cf}
\end{figure}

\figref{fig:cf} compares $S(z)$ with the result numerically
obtained from $\phi(q)$ by the Legendre transformation in the region
of $dS(z)/dz>0$. Order $q$ of the moment is increased from 0 to 8,
near and beyond which the scaling of moments, $\avg{y_t^q}\propto
t^{\phi(q)}$, becomes unclear. This might be related to an ill
definition of higher moments caused by the heavy-tail behavior of the
PDF for $y_t$ \cite{plerou99}. Near the minimum of $S(z)$ ($z<-0.3$),
the curve obtained from $\phi(q)$ can be closely approximated by a
quadratic polynomial \cite{mandelbrot97}, and fits reasonably to the
curve of $S(z)$. However, it deviates significantly from $S(z)$ for
the region of larger fluctuations, where $S''(z)$ approaches zero.
This indicates a breakdown in the relation between $\phi(q)$ and
$S(z)$, which is based on the concavity of $S(z)$.  Actually, if we
approximate $S(z)$ in the region $-0.3<z<-0.1$ by using the linear
relation, $S(z)=a(z-z_0)$, with $a\sim3$, the result corresponds to
the power-law $P_t(y)\sim y^{-4}$.  This is compatible with the
heavy-tail behavior found in \cite{plerou99}, although more detailed
study is needed to clarify the functional form $S(z)$ for this region
and those of even larger fluctuations.

The Cram\'er function $S(z)$ is related to the multifractal spectrum
investigated in \cite{mandelbrot97}. Suppose the time-series with span
$T$ is divided into $(T/t)$ intervals each of which has a time span
$t$. By evaluating the exponent $\bz(t)$, \eqref{lexp}, in each
interval, one obtains the number density $N_t(z)$ for the intervals
such that $\bz(t)$ is between $z$ and $z+dz$. The fractal
dimension $f(z)$ of the support for those intervals is defined by
\begin{equation}
  N_t(z)\sim(t/T)^{-f(z)} .
\end{equation}
Noting that $Q_t(z)\sim N_t(z)/(T/t)$ results in the relation
\begin{equation}
  S(z)=1-f(z) .
\end{equation}
Thus the multifractal spectrum $f(z)$ is essentially equivalent to the
present Cram\'er function $S(z)$. It should be mentioned, however,
that the ``dimension'' $f(z)$ can take negative values because, in
principle, $S(z)$ can take any positive value. We also stress
that the function $S(z)$ can be directly calculated from data without
resorting to the scaling of moments and the Legendre transformation as
is necessary to evaluate $f(z)$. These facts may suggest that the
concept of $S(z)$ is more useful than $f(z)$.

In summary, we have found that self-similarity of volatility at
different time-scales in the strongly correlated regime can be
characterized by direct evaluation of the Cram\'er function. The
scaling relation \eqref{scaling} is our main result, which was a
consequence of the self-similarity and large deviation theory. We have
also shown the relations of the Cram\'er function $S(z)$ to the
scaling-of-moments method and to the multifractal spectrum. The
functional form of $S(z)$ is compatible with the power-law (``cubic
law'') behavior at extremely large deviation where $S(z)$ deviates
from quadratic form.

Finally we would like to remind the reader that price fluctuates
inevitably near the critical point of the apparent equilibrium between
demand and supply \cite{sornette99}. The critical point is not
observable to any market participants whose expectations and memory
give rise to the interesting statistical features of price
fluctuations. It is indeed intriguing that the critical fluctuations
in market can be well characterized by methods in non-equilibrium
statistical physics. While we have not reached an understanding of how
market dynamics gives rise to the scaling and self-similarity found
here, we believe that our finding and analysis will provide a good
basis for phenomenology challenging any models used to understand
market dynamics.

% acknowledgement
\vspace{1em}

  Y.F. would like to thank H.~Takayasu for various information in this
  discipline, M.~Aoki for relevant concepts in economics, as well as
  S.~Maekawa, T.~Watanabe and Y.~Nakayama for discussions. This work
  is partially supported by Grant-in-Aid for Scientific Research
  No.~11837009 from the Ministry of Education, Science, Sports and
  Culture of Japan.

% --------------------------------------------------

% --------------------------------------------------
\clearpage
\thispagestyle{empty}

\clearpage
\thispagestyle{empty}

\clearpage
\thispagestyle{empty}

% --------------------------------------------------

\begin{thebibliography}{00}
\bibitem{footnote1}
  Briefly, each day in NYSE is divided into bins with an interval of 5
  minutes, for each of which an average of $y_t(s)$ is calculated for
  the entire dataset with $s$ being in the bin. Then, $y_t(s)$ is
  normalized by the average according to the bin of $s$.
\end{thebibliography}
\end{document}